%% file: main.tex
\documentclass[journal, compsoc]{IEEEtran}

\input{packages}

\begin{document}
\title{System on Chip Rejuvenation in the Wake of Persistent Attacks}

\author{\IEEEauthorblockA{\emph{(Work In Progress) \\}}
        
        \IEEEauthorblockN{Ahmad~T~Sheikh, %~\IEEEmembership{Member,~IEEE,}
        Ali~Shoker, %~\IEEEmembership{Member,~IEEE,}
        and~Paulo~Esteves-Ver{\'i}ssimo~%\IEEEmembership{Fellow,~IEEE,} 
        \\}% <-this % stops a space
        
        \IEEEauthorblockA{\emph{Resilient Computing and Cybersecurity Center (RC3), \\
        Computer, Electrical and Mathematical Sciences and Engineering Division (CEMSE), \\
        King Abdullah University of Science and Technology (KAUST) \\
        Thuwal 23955-6900, Kingdom of Saudi Arabia.}}
    
    \thanks{Ahmad T. Sheikh is Postdoctoral Fellow at RC3 KAUST. Saudi Arabia. 
    (email: \href{mailto:ahmad.sheikh@kaust.edu.sa}{ahmad.sheikh@kaust.edu.sa})}

   \thanks{Ali Shoker is Staff Research Scientist at RC3 KAUST. Saudi Arabia. 
    (email: \href{mailto:ali.shoker@kaust.edu.sa}{ali.shoker@kaust.edu.sa})}
    
    \thanks{Paulo Esteves-Verissimo is the Professor of Computer Science and Director of RC3 at KAUST. Saudi Arabia. 
    (email: \href{mailto:paulo.verissimo@kaust.edu.sa}{paulo.verissimo@kaust.edu.sa})}
}


\input{abstract}

\maketitle

\IEEEdisplaynontitleabstractindextext
\IEEEpeerreviewmaketitle

\input{introduction}
%\input{related-work}
%\input{threat-model}
\input{background}
\input{proposed-framework}
\input{feasibility}
\input{conclusion}

\bibliographystyle{IEEEtran}
\bibliography{main}

%biographies
%\input{biography}

\end{document}

%% file: packages.tex
\usepackage{cite}
\usepackage{graphicx}
\usepackage{amsmath}
\usepackage{algorithmic}
\usepackage{array}
\usepackage[hyphens]{url}
\usepackage{hyperref}
\usepackage{cite}
\usepackage{caption}

%% file: abstract.tex
\IEEEtitleabstractindextext{%
\begin{abstract}
To cope with the ever increasing threats of dynamic and adaptive persistent attacks, Fault and Intrusion Tolerance (FIT) is being studied at the hardware level to increase critical systems resilience. Based on state-machine replication, FIT is known to be effective if replicas are compromised and fail independently. This requires different ways of diversification at the software and hardware levels. In this paper, we introduce the first hardware-based rejuvenation framework, we call \textit{Samsara}, that allows for creating new computing cores (on which FIT replicas run) with diverse architectures. This is made possible by taking advantage of the programmable and reconfigurable features of MPSoC with an FPGA. A persistent attack that analyzes and exploits the vulnerability of a core will not be able to exploit it as rejuvenation to a different core architecture is made fast enough. We discuss the feasibility of this design, and we leave the empirical evaluations for future work.
\end{abstract}

}

%% file: introduction.tex
\IEEEraisesectionheading{\section{Introduction}
\label{sec:introduction}}

\IEEEPARstart{T}{here} is an ever increasing reliance on hardware accelerators including GPU, ASIC, DPU, GPGPU, and FPGA to boost the performance and security of vital digital computing embedded applications, e.g., in the Internet-of-Things services, Cyber-Physical Systems, and automation. More recently, there has been growing interest in accelerating cloud applications by deploying hundreds of cores in cloud FPGA instances~\cite{karandikar2018firesim}. While the performance gains are referred to eliminating the software stack that underlies applications and using multicore processing (e.g., as in image processing and AI applications), security is believed to be near unbeatable due to the programmable immutability in face of software attacks. This caused a leap in both, domain-specific hardware secure elements with isolation and cryptographic capabilities like enclaves, hardware secure modules, vaults, etc.~\cite{SGX2016, ARMTrustZone, kinney2006trusted, sabt2015trusted}, and in hardening modern complex embedded systems, based on smaller---and thus easily verifiable---secure abstractions~\cite{Verissimo2006, MinBFT2013}. Unfortunately, the blessing of immutability is lost with the advent of general-purpose programmable hardware computing like GPGPUs and FPGAs. This makes the corresponding systems and applications more vulnerable to intrusion attacks and vulnerabilities in the hardware design itself~\cite{RowHammer2014, Spectre2019, Meltdown2020}. This appeals for the introduction of new techniques to improve the security and resilience of these programmable hardware accelerators in face of intrusion attacks and runtime errors.

Being the last line of defense in the hardware/software stack, hardware accelerators' security, both programmable and non-programmable, have been the focus of academia and industry~\cite{fournaris2017exploiting, prinetto2020hardware}. The followed approaches are mainly: (1) computing in isolation, (2) bus compartment, (3) frequent DRAM refresh and randomization~\cite{zhang2014secure}, (4) bitstream encryption~\cite{BitStreamEncryption} etc. 
%Despite this, various vulnerabilities have been reported in RTOS’s (CWE- 119, CWE-120, CWE-126, CWE-134, CWE-398, CWE-561, CWE-563)~\cite{Boghdady2021}. 
Although often secure and dependable, these techniques fall short to defend (1) intrusion attacks being hard to detect in practice due to the huge design space of the fabric, which allows for stealthy logic, kill switches~\cite{adee2008hunt}, (2) glitches due to design mistakes or dust, aging, and overheating~\cite{merlino2004dusty, celaya2010accelerated}, and (3) vulnerabilities in RTOS’s~\cite{Boghdady2021}.
%(CWE- 119, CWE-120, CWE-126, CWE-134, CWE-398, CWE-561, CWE-563)
Furthermore, the hardware verification process is a very daunting and costly task, which makes it unaffordable at large-scale production~\cite{adee2008hunt}.

To circumvent the intrusion detection inefficiency, a recent approach is to use intrusion masking~\cite{Verissimo2003, Ines2020}, inspired from state-machine-replication (SMR)~\cite{schneider1990implementing} in Distributed Computing. To tolerate up to $t$ malicious or anomalous replicas, intrusion masking requires running a number ($n$) of concurrent replicas of a process, thus forming a \textit{replicated state machine}, running on multiple processing cores in this case. The outcome of the state machine is the agreement result of a non-compromised $n-t$ quorum of replicas (both, software and underling hardware cores). Agreement is achieved by running a variant of intrusion agreement protocols, commonly known as Byzantine Agreement (BA)~\cite{cachin2005random}. Despite resilience to (unintentional) glitches and intrusions without the need to define or know apriori, there are two noteworthy limitations in this approach. The first limitation is that replicas are assumed to be compromised or fail independently, i.e. no common vulnerabilities or glitches among the replicas exist. Failure to do so however, can lead to common mode failures of $t' > t$ replicas, and thus violate the quorum invariant (since $n-t' < n-t$). The second limitation is that these systems are fixed in size ($n$). This makes them non dynamically reactive to variable threat severity levels, where the system is prone to more than $t$ simultaneous intrusions. Indeed, a persistent adversary that is given long enough time could lead to resource exhaustion~\cite{Sousa2005} in the system, i.e., compromising more than $t$ replicas.

In this paper, we introduce the first SoC rejuvenation framework, we call \emph{Samsara}, for fault and intrusion tolerance (FIT). \textit{Samsara} makes use of the FPGA partial reconfigurable regions (PRR) features of an MPSoC to spawn new CPU cores, called \textit{tiles}, on which FIT replicas can run. This targets the two main limitations of FIT protocols: (1) It allows for hot-swappable scaling out/in the number $n$ of CPU cores (hosting the replicas), and thus adjusting the intrusion resilience of the compromised replicas $t$; and (2) it improves independence of failures and resilience to advanced persistent attacks, by the ability to spawning diverse CPU cores, made available by a library of diverse \textit{softcore} templates, e.g., offered by several vendors~\cite{makni2016comparison}.

The idea of rejuvenation is not new in FIT, as it has been suggested in~\cite{sousa2006proactive} to improve the resilience of software-based systems. Nevertheless, this concept has never been used at the hardware embedded level in a dynamic way. The reason is obvious, since the architecture of non-programmable multicore hardware is pre-defined and fixed at the manufacturing phase, thus making it impossible to diversify these cores in production. Fortunately, reconfigurable hardware like FPGA have become enabler for diversifying cores, instantiated from different vendor templates. On the other hand, although literature~\cite{Behl2014BFTMulticore, Zhenyu2014BFTMulticore} have introduced the ability to control the thread-level parallelism in multicore systems, this was not dynamic and practical as it required rebooting the entire system, contrary to \emph{Samsara} that spawns and restarts CPU cores and replicas at runtime.

In this work, we introduce a preliminary architecture (in Fig.~\ref{fig:samsara}) of \emph{Samsara} framework as part of a Xilinx (AMD) Zynq architecture~\cite{ZynqMPSoC}. Nevertheless, the proposed architecture is modularized in a generic way to be integrated into similar MPSoC architectures with FPGAs. The modularity also supports future extensions for further optimization, like diverse softcores and triggering policies, i.e., periodic, proactive, and reactive. Our future work includes an empirical evaluation to validate the behavior, security, and performance in practice.

%The rest of the paper is organized as follows...

The rest of the paper is organized as follows: Section~\ref{sec:background} discusses the background on the Fault and Intrusion Tolerance (FIT) and hardware MPSoC accelerators. The proposed framework, including threat model, architecture and workflow, are described in Section~\ref{sec:proposed-framework}. We then drive a feasibility discussion in Section~\ref{sec:feasibility}, and finally conclude in Section~\ref{sec:conclusion}.

%%%%%%%%%%%%%%%%%%%%%%%%%%%

%% file: background.tex
\section{Background}
\label{sec:background}
This section gives a gentle background about Multi-processor System on Chip (MPSoC), e.g., Zynq Architecture~\cite{Xilinx2019}, and Fault and Intrusion Tolerance (FIT) to make the understanding of the following concepts smoother.

\subsection{MPSoC Architecture}
\label{sec:zynq-mpsoc}
An MPSoc is a system on a chip (SoC) which includes multiple microprocessors, often used in embedded devices. A typical modern can MPSoC~\cite{Xilinx2019} includes multiple fabricated processing cores (called \textit{hardcore}) and a programmable hardware, called Field Programmable Gate Arrays (FPGA). The latter gives the ability to reconfigure the underlying fabric with arbitrary custom-hardware logic, even after fabrication. A common case is to spawn an emulated processing core, we call it a \textit{tile}, from a \textit{softcore} template (a \textit{bit file}). The cost of development and deployment is negligible for low-medium volumes. However, per-device cost is expensive. Application Specific Integrated Circuits (ASICs) are the chips with immutable logic circuits. They are preferred for large scale manufacturing due to cheaper per device cost. FPGAs provide a suitable middle ground before for design testing and verification before taking it to the fabrication facility for tapeout.

We overview MPSoC by demonstrating on a well known Zynq MPSoC architecture proposed by Xilinx in 2016~\cite{Xilinx2019}. Zynq MPSoC is an updated version of the Zynq-7000~\cite{Zynq7000SoC} class of devices with additional processing units and improved FPGA fabric. Other FPGA or MPSoCs variants~\cite{ZynqRFSoC, ZynqVersal} are similar to the main Zynq design, despite having some implementation differences.

\subsubsection{Zynq Architecture as a Case Study}
\label{subsec:zynqMPSoC}
\begin{figure}[t]
\centering
\includegraphics[width=.8\columnwidth]{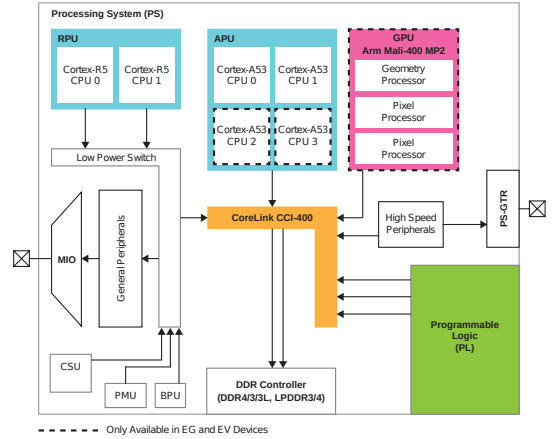}
\caption{Simplified Architecture of the Zynq Device~\cite{Xilinx2019, ExploringZynq2019}}
\label{fig:zynqMPSoC}
\end{figure}

Figure~\ref{fig:zynqMPSoC} shows a simplified architecture of the Zynq Ultrascale+ MPSoC device. The Zynq MPSoC devices consist of a Processing System (PS) and a Programmable Logic (PL) on the same chip. The PS consists of various processing elements and is responsible for managing the SoC functionality and security. One important component of PS is the Application Processing Unit (APU) that houses the Arm processing cores. These cores can be programmed using C/C++ language. PS also includes Graphics Processing Unit (GPU), Real-Time Processing Unit (RPU), interfaces to the external and internal peripherals. The Configuration Security Unit (CSU) consists of various blocks; in particular, the \emph{Processor Configuration Access Port (PCAP)} that can be used by PS to program the PL dynamically at run-time or statically.

The PL consists of re-configurable FPGA fabric that can be programmed to accelerate arbitrary any function. PL can be configured directly or by the PS. The program used to configure the PL fabric is called \texttt{bitstream} or \texttt{bitfile}. Once a PL is programmed, it becomes immutable i.e., the free and reconfigured regions of PL can't be further changed without reconfiguring the whole fabric. 
%The resources contained in the PL vary between MPSoC families. 
The PL consists of sea of Configurable Logic Block (CLB), DSP blocks, I/O pins, programmable interconnects, Block RAMs (BRAMs) etc. However, it has limited number of resources as compared to the PS.  We, therefore, aim to utilize the PL resources frugally and envisage to reconfigure them dynamically through PS as needed i.e. \emph{Partial Reconfiguration (PR)}. PR is rebranded as \emph{Dynamic Function eXchange (DFX)}~\cite{XilinxDFX} by Xilinx.

\subsubsection{Partial Reconfiguration (PR)}
\label{subsec:zynq-pr}
\begin{figure}[t]
\centering
\includegraphics[width=.6\columnwidth]{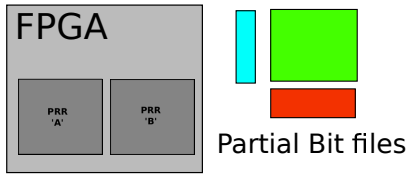}
\caption{Basic PR concept}
\label{fig:pr}
\end{figure}

%TODO: Can this be integrated to the FPGA subsection above, and then the Zynq subsection follows it as a partciular case?
FPGA technology provides an opportunity for on-site programming and reconfiguration without sending the design to the fabrication facility for modification(s). PR takes this to next level by introducing the capability to program certain regions of the PL called Partially Reconfigurable Regions (PRRs) with partial bitstreams. A full bitstream configures the FPGA device and sets it up for the acceleration, partial bitstreams can be downloaded on PL to modify only the PRRs without compromising the integrity of applications running on the remaining portions of the PL fabric. Figure~\ref{fig:pr} shows a simple example of PR, where FPGA is divided into two dynamic regions. The partial bit files are shown in different colors and using DFX they can be deployed in any dyanmic region during run time.
One of the biggest advantage of PR is the ability to time multiplex the underlying silicon for varying tasks. This would result in lesser design area and power consumption - the two most important factors of any digital design. However, accelerators orthogonal to each other can be converted to partial bit files for PR.

The PL can be programmed either through the PS or internally by the PL. From PS, the PL is programmed through the PCAP interface which transfers the full or partial bitstreams from the external DDR Memory using the DMA controller. The PL can also be programmed through the native Internal Configuration Access Port (ICAP). The ICAP has much higher speed of reconfiguration than the PCAP.

\subsection{Fault and Intrusion Tolerance}
The concept of Fault and Intrusion Tolerance (FIT)~\cite{Verissimo2003} has been proposed and studied thoroughly in the Distributed Systems area to improve system's state integrity when arbitrary faults (commonly known as Byzantine faults~\cite{BFT1982}) or intentional intrusions exist. The idea is to replicate a system process in such a way that concurrent replicas form a single \textit{deterministic state machine}~\cite{schneider1990implementing} that ensures agreement on a unique final state, i.e., using a Byzantine Agreement protocol~\cite{PBFT2002}. The correctness of FIT hinges on a quorum of correct (i.e., not faulty or malicious) nodes that have the ability to reach agreement/consensus, to maintain total order of operations. To tolerate a number $t$ of compromised replica, a total number of system replicas $n$, typically $n=3t+1$, is required~\cite{PBFT2002, zyzzyva2010}. 

The underlying assumption in FIT protocols is that replicas fail independently i.e., no common vulnerabilities among the replica nodes, failure to do so however, can lead to common mode failures. Fault and intrusion independence can be achieved by diversifying the replicas at different levels of abstraction, e.g., using N-version programming techniques~\cite{Kanoun1993, HOSSEINZADEH2018} to develop structurally different but functionally equivalent pieces of codes, or using different combinations of operating systems having non-overlapping fault vulnerabilities~\cite{lazarus2019}.

%% file: proposed-framework.tex
\section{Samsara Fault \& Intrusion Tolerance Framework}
\label{sec:proposed-framework}

\subsection{System and Threat Models}
\label{sec:threat-model}
We consider a System on Chip composed of a static processing section, called \textit{Processing System} (PS), a reconfigurable section, i.e., like an \textit{FPGA Programmable Logic} (PL), and an application-specific integrated circuit (\textit{ASIC}) chip section. All sections are connected via a reliable hardware on the same chip. For better security, \textit{Samsara's} main logic resides in the ASIC section, and is thus immutable; whereas other optimization modules reside and run in the PS. The PL section is however mutable, and used to spawn new \textit{tiles} (from softcore templates) that can host FIT replicas. The number of tiles to be spawn depends on the use of the FIT protocol~\cite{PBFT2002,MinBFT2013,Behl2014BFTMulticore}. We do not assume a particular FIT protocol in this draft.
Tiles inside the PL are connected through a reliable bus that can deliver messages to their destination. (In harsh environments, one may assume an unreliable hardware bus that may drop or modify messages exchanged, e.g., due to external factors, however, we avoid this for simplicity.) The communication between tiles could be synchronous or \textit{partially-synchronous}~\cite{PBFT1999} (i.e., deliver within an unknown bound). We assume that tiles have a decent level of containment or isolation, such that an anomaly or vulnerability may not affect other cores directly. 

In addition, we assume a strong and advanced persistent adversary that can attack the entire SoC with the aim to break the integrity of the system state (protected by the FIT protocol). Therefore, the adversary may exploit a vulnerability in the tiles (in the PL) or the above software stack (usually small operations without an OS stack). Since our focus is on hardware diversity, we assume that the system has some level of software diversity~\cite{sousa2006proactive}. Hardware diversity is however provided by \textit{Samsara} through creating diverse tiles from different softcore templates (e.g., from different vendors). However, we assume that no more than $t$ cores can be compromised during the reference time $T_a$. Within this time frame, \textit{Samsara} is expected to rejuvenate cores.
%The PS section is considered to be less vulnerable to attacks since it is not programmable, and thus immutable. 
The \textit{Samsara} modules in the PS are assumed to be encrypted and run in a secure way, e.g, in a TEE enclave (like the \textit{ARM TrustZone}~\cite{ARMTrustZone}). The aim is to protect the code from modification (both at-reset and under execution). On the other hand, Samsara's main logic is immutable being implemented in an \textit{ASIC}. This assumption is reasonable as long as we keep \textit{Samsara's} code footprint small. Finally, we assume that the adversary can compromise the cryptographic authentication keys of replicas, but cannot break the cryptographic abstractions (e.g., signatures and hash functions) using brute force.

%TODO:
\begin{figure}[t]
\centering
\includegraphics[width=0.8\columnwidth]{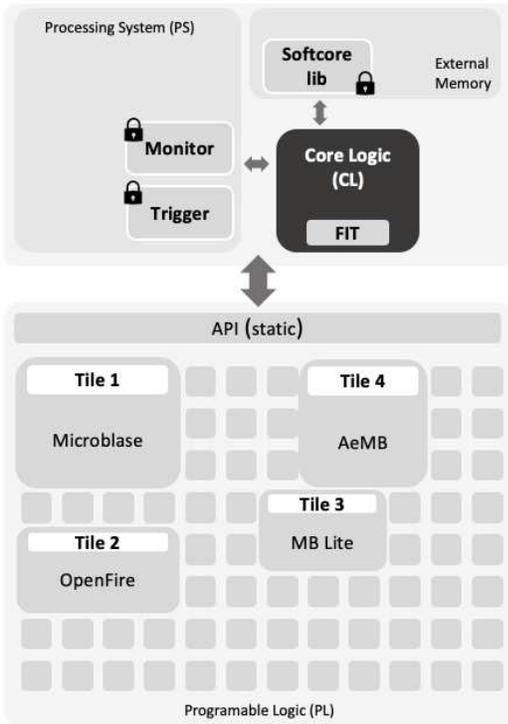}
\caption{The architecture of \textit{Samsara} Rejuvenation Fault and Intrusion Tolerant Framework on a general purpose \textit{Zynq} board architecture~\cite{ZynqMPSoC}. The four tiles are spawned from four known softcores~\cite{makni2016comparison} to demonstrate diversity.}
\label{fig:samsara}
\end{figure}

\subsection{Concept}
The idea of rejuvenation in \textit{Samsara} is based on relaunching diverse computing instances, i.e., tiles, in an FPGA, from available \textit{softcore} \textit{bit codes} templates of different internal architectures. The goal is to boost the effectiveness of hardware-based Fault and Intrusion Tolerance (FIT) in embedded systems through diversity and adaptability to different threat levels. Unfortunately, an FIT protocol is as effective as replicas are diverse, i.e., they have different implementation or design internally, but have identical specification to maintain the same functionality and behavior externally. If replicas (both hardware and software layers) are identical, there is a high probability that they err together either (1) because of hitting a common anomaly, or (2) due to a common vulnerability exploited at many replicas at once---if a persistent adversary managed to gain access to these replicas. \textit{Samsara} focuses on the rejuvenation at the hardware level. It helps instantiating diverse tiles from different templates, in a dynamic and hot-swappable way, to increase the diversity of the computing unit even under attacks. In particular, \textit{Samsara} framework has the following two main objectives:

\begin{itemize}
    \item[1.] the ability to launch a \textit{same-function-different-build} computing cores (tiles) to ensure diversity, and
    \item[2.] the ability to modify the number of tiles at runtime.
\end{itemize}

\subsection{Architecture}
We present a high-level architecture of \textit{Samsara} in Figure~\ref{fig:samsara}. Samsara is composed of three main parts: (1) the Core Logic (CL) that is an ASIC fabric; (2) the Processing System (PS) that represents a hardcore processing unit, and (3) the Programmable Logic (PL) representing an FPGA. The PS and Pl are typical in most MPSoCs, e.g., the Zynq-7000 fabric, whereas the CL is a new ASIC component to be integrated to the fabric. 

The CL controls the main rejuvenation logic and runs the FIT protocol. Being critical, the CL is proposed to be implemented on an ASIC hardware circuit fabric to be immutable, and thus immune to the attacks on the reconfigurable section (in PL). This is possible as long as the CL has a small footprint. The CL has access into the \textit{Softcore lib} that represents a set of diverse CPU core templates in the form of \texttt{bit files} stored securely in external memory. In addition, the CL has access to the \textit{Monitor} and \textit{Trigger} auxiliary modules in the PS (explained next in detail). These modules are used to extend the \textit{Samsara'a} responsiveness capabilities in a modular way.

The PL is the \textit{workplace} section hosting the created tiles, in which embedded operations are executed, and thus where rejuvenation will take place. In particular, the PL is divided into \textit{static} and \textit{dynamic} regions. The \textit{static} region holds immutable logic with API that glue PS and PL together. The \textit{dynamic} region consists of reconfigurable blocks (\texttt{pblocks}) that can be reprogrammed during run time.

\subsection{Workflow}
A typical execution workflow in \textit{Samsara} occurs as follows. An embedded application (whose state integrity is critical) invokes an operation by calling the CL. The latter runs the operation in replicated way using the implemented FIT protocol. According to the number of replicas $n$ the FIT protocol needs, the CL instantiates $n$ corresponding tiles in the PL. This is done by selecting random (and thus diverse) softcore templates from the Softcore lib. A simple FIT protocol can have the CL send each operation to different tiles for concurrent execution, and then collecting their outcomes for comparison. The agreed (matching) outcome is finally returned to the calling application. Different FIT protocols with different complexities and guarantees can be used; however, a specific FIT protocol is out of the scope of this paper.

To diversify cores, the CL selects a running tile at random to rejuvenate. The default rejuvenation frequency $\mu\in [0,T]$ is random, in order to obfuscate the pattern from the adversary. The rejuvenation protocol is inspired from software rejuvenation in~\cite{huang1995software, Silva2021}. However, we envision possible optimizations in such a hardware-based setting, that we plan to address in the future.  In a nutshell, a rejuvenation protocol includes the following steps:
\begin{itemize}
    \item spawning a new tile in the PL using a random softcore template from the \textit{Softcore lib};
    \item performing a state transfer from other tiles to initiate the new one;
    \item destroying the \textit{retired}, subject to rejuvenation; and 
    \item launching the newly created tile that replaces the retired one. 
\end{itemize}

\subsection{Optimizations}
The default policies to choose a tile to retire and a softcore to use in spawning a new tile are kept random for simplicity, and to maintain a small code footprint, and thus implemented in the CL for the ASIC fabric. However, it is possible to follow more sophisticated policies specified in the \textit{Trigger} module and with the use of the \textit{Monitor} module, both encrypted and executed in the PS section. The monitor plays the role of a \textit{watch dog}: it uses heuristics and/or other systems to detects faulty, malicious, or abnormal behaviors upon which rejuvenation is triggered. On the other side, the following triggering policies are interesting:

\begin{itemize}
 %   \item Random: targets are selected randomly. This is the default policy implmented in the CL.
    \item Periodic: targets are selected in a periodic manner.
    \item Reactive: this is an event-based policy that is triggered by the \textit{Monitor} module, e.g., if an attack or anomaly is detected.
    \item Proactive: this is the most advanced and sophisticated policy which triggers rejuvenation even before a bad event (e.g., attack) happens. This may use heuristics or AI-based intelligence. 
\end{itemize}

Another situation where the monitor can be useful is upon increasing the number $n$ of tiles in an FIT protocol, when detecting a higher threat level, i.e., $t' > t$ need to be tolerated. A higher threat can be caused by an attacker that may have access to the softcore code (sometimes open-sourced). With enough time and resources, the (persistent) attacker can possibility identify softcore vulnerabilities to be exploited when the softcore is instantiated as a tile.  In this case, \textit{Samsara} can spawn new tiles in the PL section without retiring any other tile. Again, the newly created cores are cloned from diverse templates in the Softcore lib. We will define these details in the future work.

%% file: feasibility.tex
\section{Feasibility Discussion}
\label{sec:feasibility}
Fault and Intrusion Tolerance is being increasingly used to build resilient systems, especially with the advent of \textit{Blockchain}. It was believed that FIT SMR protocols are overkill at the hardware level~\cite{Behl2014BFTMulticore}. Indeed, FIT protocols are commonly known to be computationally demanding due to (1) quadratic $(O(n^2))$ number of exchange of messages between the replicas for reaching the consensus, and (2) the extensive use of cryptography. However, with the advent of powerful hardware-based components, that allowed the use of trusted-trustworthy abstractions (i.e., hybrids)~\cite{Verissimo2006, MinBFT2013}, cryptography computation is becoming more efficient, and the spatial complexity can be reduced to $N=2f+1$, eventually requiring fairly simpler message exchange between replicas~\cite{MinBFT2013,Ines2020}. Consequently, FIT is being increasingly explored and studied in multi-core on chip embedded systems as we do here.

Similarly, rejuvenation has been proposed to diversify software as part of FIT SMR protocols~\cite{huang1995software}. However, it was not possible for FIT protocols to benefit from hardware rejuvenation~\cite{Behl2014BFTMulticore} in classical multi-core architectures. Thanks to reconfigurable/reprogrammable hardware (e.g., brought by MPSoc with FPGAs~\cite{Xilinx2019}), hardware rejuvenation became handy. Beyond that, FPGA provides a fine grain control of restarting or rejuvenating a certain core in hot-swappable fashion i.e., without the need to restart the entire system. This is subject to the availability of different core templates that are compatible with a single MPSoC platform.

One of the security challenges of \textit{Samsara} is that both the controller side (Core Logic) and the programmable side are on the same chip. This allows the adversary to compromise the controller to hamper rejuvenation, and then attack the tiles easily. For this, \textit{Samsara's} Core Logic (CL) must be highly protected, using ASIC hardware on chip implementation. Our challenge is to keep the CL small to be implemented in hardware efficiently. For this, our design separates the optimization modules (Monitor and Trigger) due to their overhead. In the worse case, the CL itself is protected and can use the default rejuvenation configurations to make the tiles more resilient to intrusion attacks.

Finally, this conceptual analysis requires an empirical experimentation to be able to validate the envisioned correctness and performance. We are currently implementing a \textit{Samsara} Proof-of-Concept prototype on a Xilinx Zynq board~\cite{Xilinx2019}. Nevertheless, our goal is to keep the \textit{Samsara} architecture generic, and thus support other SoCs. Among the interesting metrics to measure are the: code footprint, processing overhead, FIT throughput and latency, and rejuvenation time. We also aim at adjusting the protocols used and providing corresponding correctness and security proofs.

%% file: conclusion.tex
\section{Conclusion}
\label{sec:conclusion}
We have introduced \textit{Samsara}, the first SoC rejuvenation framework for Fault and Intrusion Tolerance (FIT) diversification. \textit{Samsara} leverages the programmable computational resources of an MPSoC to spawn new diverse computing cores (tiles) on which FIT replicas can run. Spawned tiles are made diverse by instantiating them from different softcore templates, likely provided by different vendors. The Core Logic of the framework is immutable and protected by hardware; whereas the tiles make use of rejuvenation to diversify the computing logic underlying the operations. We provided a conceptual analysis showing that, based on our preliminary framework, the implementation of rejuvenation on hardware is feasible. In the meanwhile, we are driving an implementation and empirical evaluation to validate our analysis in future papers.

%% file: main.bbl
\begin{thebibliography}{10}
\providecommand{\url}[1]{#1}
\csname url@samestyle\endcsname
\providecommand{\newblock}{\relax}
\providecommand{\bibinfo}[2]{#2}
\providecommand{\BIBentrySTDinterwordspacing}{\spaceskip=0pt\relax}
\providecommand{\BIBentryALTinterwordstretchfactor}{4}
\providecommand{\BIBentryALTinterwordspacing}{\spaceskip=\fontdimen2\font plus
\BIBentryALTinterwordstretchfactor\fontdimen3\font minus
  \fontdimen4\font\relax}
\providecommand{\BIBforeignlanguage}[2]{{%
\expandafter\ifx\csname l@#1\endcsname\relax
\typeout{** WARNING: IEEEtran.bst: No hyphenation pattern has been}%
\typeout{** loaded for the language `#1'. Using the pattern for}%
\typeout{** the default language instead.}%
\else
\language=\csname l@#1\endcsname
\fi
#2}}
\providecommand{\BIBdecl}{\relax}
\BIBdecl

\bibitem{karandikar2018firesim}
S.~Karandikar, H.~Mao, D.~Kim, D.~Biancolin, A.~Amid, D.~Lee, N.~Pemberton,
  E.~Amaro, C.~Schmidt, A.~Chopra \emph{et~al.}, ``Firesim: Fpga-accelerated
  cycle-exact scale-out system simulation in the public cloud,'' in \emph{2018
  ACM/IEEE 45th Annual International Symposium on Computer Architecture
  (ISCA)}.\hskip 1em plus 0.5em minus 0.4em\relax IEEE, 2018, pp. 29--42.

\bibitem{SGX2016}
\BIBentryALTinterwordspacing
V.~Costan and S.~Devadas, ``Intel sgx explained,'' \emph{Tech. rep.,
  Massachusetts Institute of Technology}, 2016. [Online]. Available:
  \url{https://eprint.iacr.org/2016/086.pdf}
\BIBentrySTDinterwordspacing

\bibitem{ARMTrustZone}
\BIBentryALTinterwordspacing
ARM, ``Armtrustzone,'' \emph{ARM}, Accessed on August 2022. [Online].
  Available: \url{https://www.arm.com/products/security-on-arm/trustzone}
\BIBentrySTDinterwordspacing

\bibitem{kinney2006trusted}
S.~L. Kinney, \emph{Trusted platform module basics: using TPM in embedded
  systems}.\hskip 1em plus 0.5em minus 0.4em\relax Elsevier, 2006.

\bibitem{sabt2015trusted}
M.~Sabt, M.~Achemlal, and A.~Bouabdallah, ``Trusted execution environment: what
  it is, and what it is not,'' in \emph{2015 IEEE Trustcom/BigDataSE/ISPA},
  vol.~1.\hskip 1em plus 0.5em minus 0.4em\relax IEEE, 2015, pp. 57--64.

\bibitem{Verissimo2006}
\BIBentryALTinterwordspacing
P.~E. Ver\'{\i}ssimo, ``Travelling through wormholes: A new look at distributed
  systems models,'' \emph{SIGACT News}, vol.~37, no.~1, p. 66–81, mar 2006.
  [Online]. Available: \url{https://doi.org/10.1145/1122480.1122497}
\BIBentrySTDinterwordspacing

\bibitem{MinBFT2013}
G.~S. Veronese, M.~Correia, A.~N. Bessani, L.~C. Lung, and P.~Verissimo,
  ``Efficient byzantine fault-tolerance,'' \emph{IEEE Transactions on
  Computers}, vol.~62, no.~1, pp. 16--30, 2013.

\bibitem{RowHammer2014}
Y.~Kim, R.~Daly, J.~Kim, C.~Fallin, J.~H. Lee, D.~Lee, C.~Wilkerson, K.~Lai,
  and O.~Mutlu, ``Flipping bits in memory without accessing them: An
  experimental study of dram disturbance errors,'' in \emph{2014 ACM/IEEE 41st
  International Symposium on Computer Architecture (ISCA)}, 2014, pp. 361--372.

\bibitem{Spectre2019}
P.~Kocher, J.~Horn, A.~Fogh, D.~Genkin, D.~Gruss, W.~Haas, M.~Hamburg, M.~Lipp,
  S.~Mangard, T.~Prescher, M.~Schwarz, and Y.~Yarom, ``Spectre attacks:
  Exploiting speculative execution,'' in \emph{2019 IEEE Symposium on Security
  and Privacy (SP)}, 2019, pp. 1--19.

\bibitem{Meltdown2020}
\BIBentryALTinterwordspacing
M.~Lipp, M.~Schwarz, D.~Gruss, T.~Prescher, W.~Haas, J.~Horn, S.~Mangard,
  P.~Kocher, D.~Genkin, Y.~Yarom, M.~Hamburg, and R.~Strackx, ``Meltdown:
  Reading kernel memory from user space,'' \emph{Commun. ACM}, vol.~63, no.~6,
  p. 46–56, may 2020. [Online]. Available:
  \url{https://doi.org/10.1145/3357033}
\BIBentrySTDinterwordspacing

\bibitem{fournaris2017exploiting}
A.~P. Fournaris, L.~Pocero~Fraile, and O.~Koufopavlou, ``Exploiting hardware
  vulnerabilities to attack embedded system devices: A survey of potent
  microarchitectural attacks,'' \emph{Electronics}, vol.~6, no.~3, p.~52, 2017.

\bibitem{prinetto2020hardware}
P.~Prinetto and G.~Roascio, ``Hardware security, vulnerabilities, and attacks:
  A comprehensive taxonomy.'' in \emph{ITASEC}, 2020, pp. 177--189.

\bibitem{zhang2014secure}
T.~Zhang and R.~B. Lee, ``Secure cache modeling for measuring side-channel
  leakage,'' \emph{Technical Report, Princeton University}, 2014.

\bibitem{BitStreamEncryption}
\BIBentryALTinterwordspacing
2022, ``Using encryption and authentication to secure an ultrascale/ultrascale+
  fpga bitstream,'' \emph{Xilinx}, 2022. [Online]. Available:
  \url{https://www.xilinx.com/content/dam/xilinx/support/documents/application_notes/xapp1267-encryp-efuse-program.pdf}
\BIBentrySTDinterwordspacing

\bibitem{adee2008hunt}
S.~Adee, ``The hunt for the kill switch,'' \emph{iEEE SpEctrum}, vol.~45,
  no.~5, pp. 34--39, 2008.

\bibitem{merlino2004dusty}
R.~L. Merlino and J.~A. Goree, ``Dusty plasmas in the laboratory, industry, and
  space,'' \emph{PHYSICS TODAY.}, vol.~57, no.~7, pp. 32--39, 2004.

\bibitem{celaya2010accelerated}
J.~R. Celaya, P.~Wysocki, V.~Vashchenko, S.~Saha, and K.~Goebel, ``Accelerated
  aging system for prognostics of power semiconductor devices,'' in \emph{2010
  Ieee Autotestcon}.\hskip 1em plus 0.5em minus 0.4em\relax IEEE, 2010, pp.
  1--6.

\bibitem{Boghdady2021}
\BIBentryALTinterwordspacing
A.~Al-Boghdady, K.~Wassif, and M.~El-Ramly, ``The presence, trends, and causes
  of security vulnerabilities in operating systems of iot’s low-end
  devices,'' \emph{Sensors}, vol.~21, no.~7, 2021. [Online]. Available:
  \url{https://www.mdpi.com/1424-8220/21/7/2329}
\BIBentrySTDinterwordspacing

\bibitem{Verissimo2003}
P.~E. Ver{\'i}ssimo, N.~F. Neves, and M.~P. Correia, ``Intrusion-tolerant
  architectures: Concepts and design,'' in \emph{Architecting Dependable
  Systems}, R.~de~Lemos, C.~Gacek, and A.~Romanovsky, Eds.\hskip 1em plus 0.5em
  minus 0.4em\relax Berlin, Heidelberg: Springer Berlin Heidelberg, 2003, pp.
  3--36.

\bibitem{Ines2020}
\BIBentryALTinterwordspacing
I.~P. Gouveia, M.~Völp, and P.~Esteves-Verissimo, ``Behind the last line of
  defense -- surviving soc faults and intrusions,'' 2020. [Online]. Available:
  \url{https://arxiv.org/abs/2005.04096}
\BIBentrySTDinterwordspacing

\bibitem{schneider1990implementing}
F.~B. Schneider, ``Implementing fault-tolerant services using the state machine
  approach: A tutorial,'' \emph{ACM Computing Surveys (CSUR)}, vol.~22, no.~4,
  pp. 299--319, 1990.

\bibitem{cachin2005random}
C.~Cachin, K.~Kursawe, and V.~Shoup, ``Random oracles in constantinople:
  Practical asynchronous byzantine agreement using cryptography,''
  \emph{Journal of Cryptology}, vol.~18, no.~3, pp. 219--246, 2005.

\bibitem{Sousa2005}
P.~Sousa, N.~Neves, and P.~Verissimo, ``How resilient are distributed f
  fault/intrusion-tolerant systems?'' in \emph{2005 International Conference on
  Dependable Systems and Networks (DSN'05)}, 2005, pp. 98--107.

\bibitem{makni2016comparison}
M.~Makni, M.~Baklouti, S.~Niar, M.~W. Jmal, and M.~Abid, ``A comparison and
  performance evaluation of fpga soft-cores for embedded multi-core systems,''
  in \emph{2016 11th International Design \& Test Symposium (IDT)}.\hskip 1em
  plus 0.5em minus 0.4em\relax IEEE, 2016, pp. 154--159.

\bibitem{sousa2006proactive}
P.~Sousa, N.~F. Neves, and P.~Ver{\'\i}ssimo, ``Proactive resilience through
  architectural hybridization,'' in \emph{Proceedings of the 2006 ACM Symposium
  on Applied Computing}, 2006, pp. 686--690.

\bibitem{Behl2014BFTMulticore}
J.~Behl, T.~Distler, and R.~Kapitza, ``Scalable bft for multi-cores:
  Actor-based decomposition and consensus-oriented parallelization,'' in
  \emph{Proceedings of the 10th USENIX Conference on Hot Topics in System
  Dependability}, ser. HotDep'14.\hskip 1em plus 0.5em minus 0.4em\relax USA:
  USENIX Association, 2014, p.~9.

\bibitem{Zhenyu2014BFTMulticore}
\BIBentryALTinterwordspacing
Z.~Guo, C.~Hong, M.~Yang, D.~Zhou, L.~Zhou, and L.~Zhuang, ``Rex: Replication
  at the speed of multi-core,'' in \emph{Proceedings of the Ninth European
  Conference on Computer Systems}, ser. EuroSys '14.\hskip 1em plus 0.5em minus
  0.4em\relax New York, NY, USA: Association for Computing Machinery, 2014.
  [Online]. Available: \url{https://doi.org/10.1145/2592798.2592800}
\BIBentrySTDinterwordspacing

\bibitem{ZynqMPSoC}
\BIBentryALTinterwordspacing
2020, ``Zynq ultrascale+ device technical reference manual,'' \emph{Xilinx},
  2020. [Online]. Available:
  \url{https://docs.xilinx.com/v/u/en-US/ug1085-zynq-ultrascale-trm}
\BIBentrySTDinterwordspacing

\bibitem{Xilinx2019}
Xilinx2019, ``Ug1085: Zynq ultrascale+ device technical reference manual,''
  \emph{Xilinx}, 2019.

\bibitem{Zynq7000SoC}
\BIBentryALTinterwordspacing
2021, ``Zynq-7000 soc technical reference manual,'' \emph{Xilinx}, 2021.
  [Online]. Available:
  \url{https://docs.xilinx.com/v/u/en-US/ug585-Zynq-7000-TRM}
\BIBentrySTDinterwordspacing

\bibitem{ZynqRFSoC}
\BIBentryALTinterwordspacing
2022, ``Zynq ultrascale+ device technical reference manual,'' \emph{Xilinx},
  2022. [Online]. Available:
  \url{https://www.xilinx.com/content/dam/xilinx/publications/product-briefs/xilinx-rfsoc-product-brief.pdf}
\BIBentrySTDinterwordspacing

\bibitem{ZynqVersal}
\BIBentryALTinterwordspacing
------, ``Versal acap technical reference manual (am011,'' \emph{Xilinx}, 2022.
  [Online]. Available:
  \url{https://docs.xilinx.com/r/en-US/am011-versal-acap-trm/Introduction}
\BIBentrySTDinterwordspacing

\bibitem{ExploringZynq2019}
\BIBentryALTinterwordspacing
L.~Crockett, D.~Northcote, C.~Ramsay, F.~Robinson, and R.~Stewart,
  \emph{Exploring Zynq MPSoC : With PYNQ and Machine Learning
  Applications}.\hskip 1em plus 0.5em minus 0.4em\relax Glasgow: Strathclyde
  Academic Media, April 2019. [Online]. Available:
  \url{https://www.zynq-mpsoc-book.com/}
\BIBentrySTDinterwordspacing

\bibitem{XilinxDFX}
\BIBentryALTinterwordspacing
Xilinx, ``Ug909: Vivado design suite user guide: Dynamic function exchange,''
  \emph{Xilinx}, 2018. [Online]. Available:
  \url{https://docs.xilinx.com/r/2021.1-English/Vivado-Design-Suite-User-Guide-Dynamic-Function-eXchange-UG909}
\BIBentrySTDinterwordspacing

\bibitem{BFT1982}
\BIBentryALTinterwordspacing
L.~Lamport, R.~Shostak, and M.~Pease, ``The byzantine generals problem,''
  \emph{ACM Trans. Program. Lang. Syst.}, vol.~4, no.~3, p. 382–401, jul
  1982. [Online]. Available: \url{https://doi.org/10.1145/357172.357176}
\BIBentrySTDinterwordspacing

\bibitem{PBFT2002}
\BIBentryALTinterwordspacing
M.~Castro and B.~Liskov, ``Practical byzantine fault tolerance and proactive
  recovery,'' \emph{ACM Trans. Comput. Syst.}, vol.~20, no.~4, p. 398–461,
  nov 2002. [Online]. Available: \url{https://doi.org/10.1145/571637.571640}
\BIBentrySTDinterwordspacing

\bibitem{zyzzyva2010}
\BIBentryALTinterwordspacing
R.~Kotla, L.~Alvisi, M.~Dahlin, A.~Clement, and E.~Wong, ``Zyzzyva: Speculative
  byzantine fault tolerance,'' \emph{ACM Trans. Comput. Syst.}, vol.~27, no.~4,
  jan 2010. [Online]. Available: \url{https://doi.org/10.1145/1658357.1658358}
\BIBentrySTDinterwordspacing

\bibitem{Kanoun1993}
K.~Kanoun, M.~Kaaniche, C.~Beounes, J.-C. Laprie, and J.~Arlat, ``Reliability
  growth of fault-tolerant software,'' \emph{IEEE Transactions on Reliability},
  vol.~42, no.~2, pp. 205--219, 1993.

\bibitem{HOSSEINZADEH2018}
\BIBentryALTinterwordspacing
S.~Hosseinzadeh, S.~Rauti, S.~Laurén, J.-M. Mäkelä, J.~Holvitie,
  S.~Hyrynsalmi, and V.~Leppänen, ``Diversification and obfuscation techniques
  for software security: A systematic literature review,'' \emph{Information
  and Software Technology}, vol. 104, pp. 72--93, 2018. [Online]. Available:
  \url{https://www.sciencedirect.com/science/article/pii/S0950584918301484}
\BIBentrySTDinterwordspacing

\bibitem{lazarus2019}
\BIBentryALTinterwordspacing
M.~Garcia, A.~Bessani, and N.~Neves, ``Lazarus: Automatic management of
  diversity in bft systems,'' in \emph{Proceedings of the 20th International
  Middleware Conference}, ser. Middleware '19.\hskip 1em plus 0.5em minus
  0.4em\relax New York, NY, USA: Association for Computing Machinery, 2019, p.
  241–254. [Online]. Available: \url{https://doi.org/10.1145/3361525.3361550}
\BIBentrySTDinterwordspacing

\bibitem{PBFT1999}
M.~Castro and B.~Liskov, ``Practical byzantine fault tolerance,'' in
  \emph{Proceedings of the Third Symposium on Operating Systems Design and
  Implementation}, ser. OSDI '99.\hskip 1em plus 0.5em minus 0.4em\relax USA:
  USENIX Association, 1999, p. 173–186.

\bibitem{huang1995software}
Y.~Huang, C.~Kintala, N.~Kolettis, and N.~D. Fulton, ``Software rejuvenation:
  Analysis, module and applications,'' in \emph{Twenty-fifth international
  symposium on fault-tolerant computing. Digest of papers}.\hskip 1em plus
  0.5em minus 0.4em\relax IEEE, 1995, pp. 381--390.

\bibitem{Silva2021}
\BIBentryALTinterwordspacing
D.~Silva, R.~Graczyk, J.~Decouchant, M.~Volp, and P.~Esteves-Verissimo,
  ``Threat adaptive byzantine fault tolerant state-machine replication,'' in
  \emph{2021 40th International Symposium on Reliable Distributed Systems
  (SRDS)}.\hskip 1em plus 0.5em minus 0.4em\relax Los Alamitos, CA, USA: IEEE
  Computer Society, sep 2021, pp. 78--87. [Online]. Available:
  \url{https://doi.ieeecomputersociety.org/10.1109/SRDS53918.2021.00017}
\BIBentrySTDinterwordspacing

\end{thebibliography}
